\documentclass[twocolumn,showpacs,preprintnumbers,amsmath,amssymb]{revtex4}

\usepackage{graphicx} 
\usepackage{dcolumn} 
\usepackage{bm}
\usepackage{txfonts}
\usepackage{mathrsfs}
\usepackage{comment}

\def\vq{{\bf q}}

\def\vk{{\bf k}}

\def\vS{{\bf S}}
\newcommand{\eq}[1]{Eq.~(\ref{#1})}
\newcommand{\fig}[1]{Fig.~\ref{#1}}
\newcommand{\be}{\begin{equation}}
\newcommand{\ee}{\end{equation}}
\newcommand{\bea}{\begin{eqnarray}}

\newcommand{\eea}{\end{eqnarray}}

\begin{document}

\title{Singular non-ordering susceptibility at a Pomeranchuk instability} 

\author{Hiroyuki Yamase$^{1}$ and Pawel Jakubczyk$^{2}$}
\affiliation{$^{1}$National Institute for Materials Science, Tsukuba 305-0047, Japan\\
$^{2}$Institute for Theoretical Physics, Warsaw University, Ho\.za 69, 00-681 Warsaw, Poland}

\date{\today}

\begin{abstract}
We study magnetic susceptibilities of two-dimensional itinerant electron systems 
exhibiting symmetry-breaking Fermi surface distortions, the so-called 
$d$-wave Pomeranchuk instability, in a magnetic field. 
In a pure forward scattering model, 
the longitudinal susceptibility $\chi^{zz}$ is found to exhibit a jump at a critical point. 
The magnitude of this jump diverges at a tricritical point. 
When scattering processes involving finite momentum transfers are allowed for, $\chi^{zz}$ is expected to diverge also at a critical point. 
The system displays multiple critical fluctuations. 
We argue that the features of $\chi^{zz}$ are general properties  
associated with singularities of a non-ordering susceptibility, 
leading to implications for a variety of materials including 
$\textrm{Sr}_3\textrm{Ru}_2\textrm{O}_7$. 
\end{abstract}

\pacs{75.40.-s, 71.18.+y, 05.50.+q, 64.60.F-} 

\maketitle

\section{Introduction}
The electronic nematic 
is currently among the most intriguing and intensely investigated states of quantum
matter. Unlike its classical counterpart \cite{deGennes93} 
this phase originates from interactions
between structureless, point-like particles (electrons). On the other
hand, just like in the case of classical nematics, these systems exhibit
orientational symmetry breaking while preserving translational invariance. 
The electronic nematic state is reached via a phase transition from either 
the electronic smectic state corresponding to the so-called charge-stripe order \cite{kivelson98} 
or the uniform state as found in the $t$-$J$ \cite{yamase00} and 
Hubbard \cite{metzner00,valenzuela01} 
models. The latter route is accompanied by spontaneous Fermi surface symmetry breaking 
with a $d$-wave order parameter, and is often referred to as 
the $d$-wave Pomeranchuk instability \cite{pomeranchuk58}. 

The occurrence of the electronic nematic was indicated in a number of 
correlated electron systems. 
In two-dimensional electron gases, the electronic nematic 
is signaled by the strong anisotropy of magnetotransport  at low temperature 
in half-filled higher Landau levels \cite{wexler06}. 
For cuprate superconductors, 
there exists a strong tendency toward the nematic state 
despite competition with cooper pairing formation, as seen 
in the pronounced anisotropy of magnetic excitation spectra \cite{hinkov08}. 
The bilayer ruthenate Sr$_{3}$Ru$_{2}$O$_{7}$ displays the evidence 
of the nematic state \cite{borzi07,rost09}, 
which is realized in a certain range of a magnetic field bounded by 
first order transition lines  
at low temperature and a second order transition line at high temperature; 
the end points of the second order transition line are tricritical points \cite{grigera04}.

Motivated by experiments performed for Sr$_{3}$Ru$_{2}$O$_{7}$, we consider the nematic instability 
from a uniform state, namely the Pomeranchuk instability (PI),  in the presence of a magnetic field. 
We find that the PI is accompanied by peculiar singularities of the uniform spin susceptibility, 
which can be observed in real materials like Sr$_{3}$Ru$_{2}$O$_{7}$. 

Our results are not specific to the PI and can be understood as universal  
singular behavior of a non-ordering susceptibility. 
The concept of a non-ordering susceptibility appears in the theory of 
tricriticality developed for helium mixtures and some metamagnetic 
materials \cite{lawrie84,stryjewski77}. 
At the tricritical point  three different phases 
become identical \cite{griffiths7073} and the full phase diagram 
is described by temperature ($T$), a control parameter called a non-ordering field, 
and an ordering field which is conjugate to the order parameter \cite{riedel72}. 
The framework naturally involves two different kinds of susceptibilities, 
the ordering and non-ordering ones. 
The former is given by the second derivative of the free energy with respect to 
the ordering field, 
whereas the latter is the second derivative with respect to the non-ordering field; 
the non-ordering susceptibility is thus defined uniquely in the system in question. 
In the context of Sr$_{3}$Ru$_{2}$O$_{7}$, the ordering susceptibility corresponds to 
the $d$-wave charge compressibility \cite{yamase05} and the non-ordering one 
to the longitudinal magnetic susceptibility.

In this paper, we first study the uniform magnetic susceptibilities 
in a mean-field model of the PI on a square lattice in the presence of a magnetic field. 
We show that the longitudinal susceptibility $\chi^{zz}$ exhibits a jump at 
a critical point (CP) and a power-law divergence at a tricritical point (TCP), 
while the transverse susceptibility $\chi^{\pm}$ exhibits a cusp. 
In two-dimensional systems, the mean-field theory breaks down in a critical regime. 
We then argue that beyond the mean-field theory 
$\chi^{zz}$ diverges at both CP and TCP, 
which is associated with multiple critical fluctuations, 
{\it i.e.}, soft Fermi surface fluctuations and magnetic fluctuations. 
The singular behavior of $\chi^{zz}$ is interpreted as a general feature 
associated with singularities of a non-ordering susceptibility 
at a phase transition, leading to 
implications for  a variety of materials including Sr$_{3}$Ru$_{2}$O$_{7}$.

\section{Model and formalism}
\begin{figure}[ht!]
\begin{center}
\includegraphics[width=6.5cm]{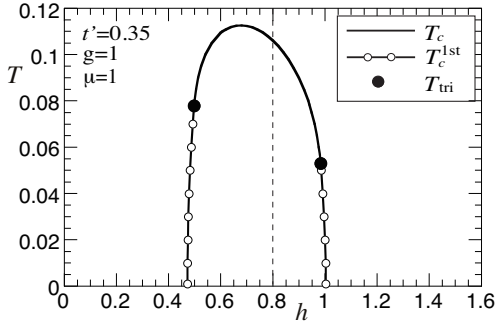}
\caption{The phase diagram in the plane of $h$ and $T$ 
for $t'=0.35$, $g=1$, and $\mu=1$.  
$T_{c}$ ($T_{c}^{\rm 1st}$) denotes a second (first) order 
transition and $T_{\rm tri}$ is a tricritical point. The dashed line represents the 
position of the van Hove energy.} 
\label{phase}
\end{center}
\end{figure}

We consider the following Hamiltonian for the PI \cite{yamase07b,yamase07c} 
\begin{equation}
\label{f-model}
H=\sum_{\vk,\sigma}
(\epsilon_\vk^0-\mu)n_{\vk\sigma}+\frac{1}{2N}\sum_{\vk,\vk',\sigma,\sigma'}
f_{\vk\,\vk'} n_{\vk\sigma}n_{\vk'\sigma'}-
\frac{h}{2}\sum_{\vk,\sigma}\sigma n_{\vk\sigma}\; ,
\end{equation}
describing tight-binding electrons on a square lattice, interacting via a
forward scattering interaction and subject to an external magnetic
field $h$. The dispersion $\epsilon_\vk^0=-2[t(\cos k_x +\cos k_y) +2t' \cos k_x
\cos k_y] $ retains the nearest and next-nearest neighbor hopping amplitudes; 
$\mu$ is the chemical potential; $n_{\vk\sigma}$ is the electron number operator 
with momentum $\vk$ and spin $\sigma$; 
$N$ is the total number of lattice sites; the effective interaction 
\begin{equation} 
f_{\vk\,\vk'}= -g d_{\vk} d_{\vk'}
\label{fkk}
\end{equation}
involves the $d$-wave form factor $d_\vk=\cos k_x -\cos k_y$ and 
drives the PI. This interaction mimics the actual structure of the effective charge interactions 
in the forward scattering channel as obtained 
within more fundamental models \cite{yamase00,metzner00,valenzuela01}. 
The order parameter corresponding to the PI is given by 
\begin{equation}
\label{ord_par_def}
\eta=-\frac{g}{N}\sum_{\vk,\sigma} d_{\vk} \left< n_{\vk\sigma} \right> \;.
\end{equation}
The transition can be tuned along an isotherm either by varying the chemical potential $\mu$ or the
magnetic field $h$ in the Hamiltonian (\ref{f-model}). 
The latter mimics the actual experimental
setup in $\textrm{Sr}_3\textrm{Ru}_2\textrm{O}_7$. 
In fact, the model (1) and different versions of it 
turned out to capture numerous aspects possibly associated with 
the $\textrm{Sr}_3\textrm{Ru}_2\textrm{O}_7$ compound  \cite{kee05,doh07,puetter07,doh07b,yamase07b,yamase07c,ho08,yamase09b,adachi09,fischer10}.  
We fix the band parameters $t'=0.35$, $g=1$, 
and $\mu=1$ in units set by $t$ \cite{yamase07b,yamase07c}. 
The theoretical phase diagram is shown in \fig{phase}.   
The PI is realized in a region around the van Hove energy ($h=0.8$) of the 
up-spin band. 
A second-order phase transition occurs at high $T$, but terminates at a TCP, 
below which the transition becomes first order. 

\begin{figure}
\begin{center}
\includegraphics[width=7cm]{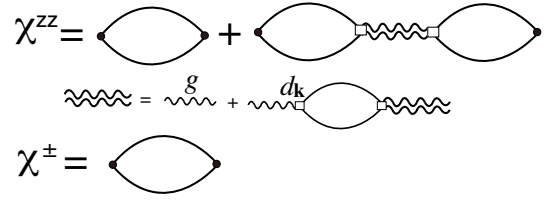}
\caption{Graphical representation of $\chi^{zz}$ and $\chi^{\pm}$ for the Hamiltonian (\ref{f-model}).   
The vertex with a circle (square) indicates the form factor 1 ($d_{\vk}$). 
The solid line denotes electron propagator with the dispersion $\xi_{\vk\sigma}$. 
}
\label{diagram}
\end{center}
\end{figure}

We now analyze the uniform longitudinal magnetic susceptibility $\chi^{zz}$ 
within the model (\ref{f-model}). 
The standard linear response theory yields 
\be
\chi^{zz}= \lim_{\vq \to 0} \frac{i}{N} \int_{0}^{\infty} {\rm d}t 
{\rm e}^{-\delta t} \langle 
[S^{z}(\vq,\,t), S^{z}(-\vq,\,0)]
\rangle \;.
\ee
Here $\langle  \cdots \rangle$ denotes the equilibrium
 expectation value with the Hamiltonian (\ref{f-model}),  
$[\cdot,\cdot]$ is the commutator; $\delta$ a positive infinitesimal; 
$\vS (\vq,t) ={\rm e}^{{\rm i}Ht} \vS(\vq) {\rm e}^{-{\rm i}Ht}$ and $ \vS(\vq) $ is the 
spin operator in momentum space. 
In the random phase approximation (RPA), 
$\chi^{zz}$ is given graphically by \fig{diagram} and calculated as 
\begin{equation}
\label{chizz}
\chi^{zz}=\frac{1}{4}\left[-\frac{1}{N}\sum_{\vk,\sigma}f'(\xi_{\vk\sigma})
+\frac{g\left(\frac{1}{N}\sum_{\vk,\sigma}\sigma
      d_{\vk}f'(\xi_{\vk\sigma})\right)^2}{1
      +\frac{g}{N}\sum_{\vk,\sigma}d_{\vk}^2f'(\xi_{\vk\sigma})}\right]\;,
\end{equation}
where $\xi_{\vk\sigma}=\epsilon_{\vk}^0-\mu+\eta d_{\vk}-\frac{\sigma}{2}h$; 
$f(x)$ is the Fermi-Dirac distribution and its first derivative is denoted by  $f'$. 
The presence of the second term in \eq{chizz} is crucial to the present study 
and was overlooked in Ref.~\onlinecite{yamase07b}. 
On the other hand, the uniform transverse susceptibility is expressed 
by a simple bubble (\fig{diagram}) in the RPA, that is, 
\bea
&&\chi^{\pm}=\lim_{\vq \to 0} \frac{i}{N} \int_{0}^{\infty} {\rm d}t 
{\rm e}^{-\delta t} \langle 
[S^{+}(\vq,\,t), S^{-}(-\vq,\,0)]
\rangle \\
&& \hspace{4.5mm} =-\frac{1}{N}\sum_{\vk}\frac{f(\xi_{\vk \uparrow})-
f(\xi_{\vk \downarrow})}{\xi_{\vk \uparrow}-\xi_{\vk \downarrow}} \\
&& \hspace{4.5mm} =\frac{2m}{h}\;, 
\label{chipm}
\eea
where $S^{\pm}=S_{x} \pm {\rm i}S_{y}$ and 
$m=\frac{1}{2N}\sum_{\vk,\sigma} \sigma f(\xi_{\vk \sigma})$ 
is the magnetization. 

\begin{figure*}[ht!]
\begin{center}
\includegraphics[width=12.5cm]{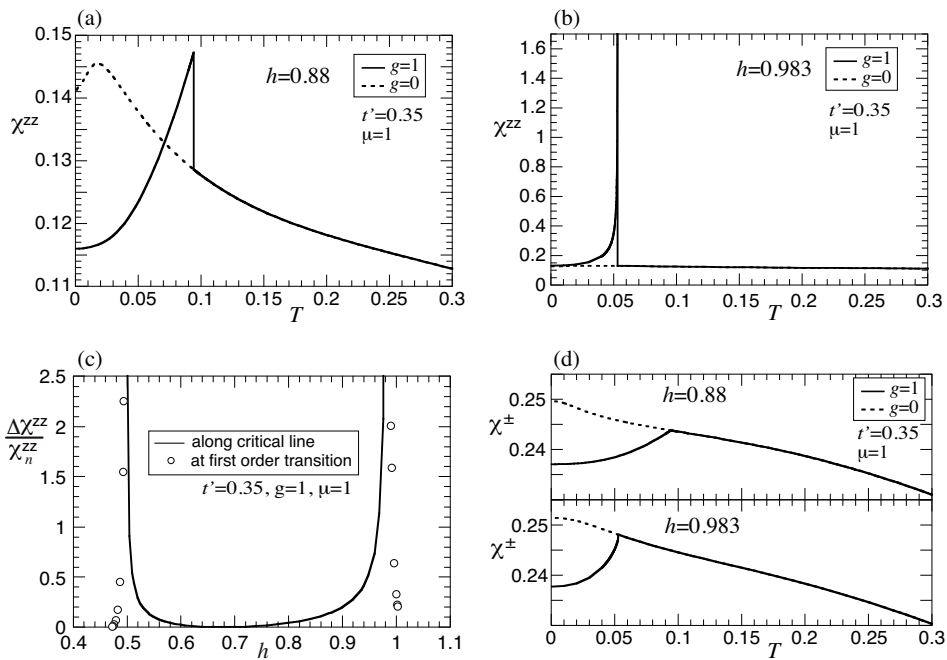}
\caption{(a) and (b) The longitudinal susceptibility $\chi^{zz}$ plotted versus 
temperature as the second order transition is approached at $h=0.88$ (a) and 
at the tricritical field $h_{\rm tri}=0.983$ (b). 
The dashed line is the result for g=0. 
(c) The jump of $\chi^{zz}$ ($\Delta\chi^{zz}$) along a second 
order transition line (solid line) and at a first order transition (circle) 
on the side of the symmetry-broken phase. 
The magnitude of the jump is scaled by $\chi^{zz}$ on the side of 
the normal phase ($\chi_{n}^{zz}$). 
(d) $\chi^{\pm}$ is plotted for $h=0.88$ and $0.983$ 
in a similar fashion to (a). 
} 
\label{MFresult}
\end{center}
\end{figure*}

\section{Results}
We plot the susceptibility $\chi^{zz}$ versus temperature in \fig{MFresult}(a) 
as the second-order transition line is crossed at fixed $h$. 
The quantity $\chi^{zz}$ exhibits a discontinuity at $T=T_c(h)$. The magnitude of
the jump diverges as the field $h$ is varied toward the tricritical value
$h_{\rm tri}=0.983$ [\fig{MFresult}(b)]. It follows the power law 
$\chi^{zz}\propto |T-T_{\rm tri}|^{-1/2} $ \cite{miscchizz}. 
The exponent $1/2$ is visible 
only sufficiently close to the TCP and very high numerical accuracy is required. 
The divergence of $\chi^{zz}$ is observed only in the symmetry-broken phase, while 
in the disordered phase it remains finite and smooth.    
In \fig{MFresult}(c), the jump of $\chi^{zz}$ ($\Delta\chi^{zz}$) is plotted 
along the transition line (solid line).  
The magnitude of the jump is strongly suppressed around $h$ where $T_{c}$ 
becomes maximal (see also \fig{phase}) 
and is gradually enhanced away from it. 
While $\chi^{zz}$ necessarily shows a jump upon crossing a first order transition line, 
we also plot its magnitude in \fig{MFresult}(c) with open circles. 
It is interesting to note that the magnitude of the jump of $\chi^{zz}$ at a CP is typically 
comparable to that at a first order transition.  
In contrast to the results for $\chi^{zz}$, 
the transverse susceptibility $\chi^{\pm}$ [\fig{MFresult}(d)]  
exhibits a cusp upon crossing the $T_c$ line 
at both CP and TCP. 
Although $\chi^{\pm}$ can be enhanced below $T_c$ for a different choice of parameters, 
its non-analyticity at $T_c$ is always a cusp.

As shown in \fig{diagram}, $\chi^{zz}$ involves contributions from two diagrams. Its peculiar 
behavior originates from the second diagram, which 
is evaluated as the second term in \eq{chizz}. 
This term can become finite only in the symmetry-broken phase 
because of the $d$-wave form factor in the numerator. 
Hence the singular behavior of $\chi^{zz}$ is found 
only in the symmetry-broken phase [\fig{MFresult}(a) and (b)]. 
This feature was observed also in different systems involving non-ordering
susceptibilities considered in the context of helium mixtures \cite{blume71}.  
On the other hand, the forward scattering interaction does not 
contribute to $\chi^{\pm}$ (\fig{diagram}), 
yielding behavior very distinct from $\chi^{zz}$ [\fig{MFresult}(d)].

\subsection{Comparison with a jump of the specific heat} 
\begin{figure}[th!]
\begin{center}
\includegraphics[width=6.5cm]{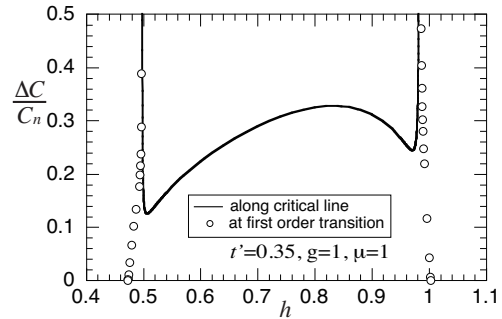}
\caption{The jump of $C$ ($\Delta\!\!~C$) along a second 
order transition line (solid line) and at a first order transition (circle) 
on the side of the symmetry-broken phase. 
The magnitude of the jump is scaled by $C$ on the side of 
the normal phase ($C_{n}$). 
} 
\label{gamma}
\end{center}
\end{figure}

It is instructive to compare the jump of $\chi^{zz}$ with 
a jump of the specific heat coefficient ($C/T$) \cite{yamase07b}, which 
is plotted along the transition line with  
a solid line in \fig{gamma}. The jump is pronounced around the van Hove 
energy ($h=0.8$) and is suppressed monotonously away from it. 
The quantity $C/T$ rapidly increases and ultimately diverges upon approaching 
the vicinity of the TCP; the singularity is given by 
$C/T\sim |T-T_{\rm tri}|^{-1/2} $ at $h=h_{\rm tri}$ or 
$C/T \sim  |h-h_{\rm tri}|^{-1/2}$ at $T=T_{\rm tri}$ sufficiently close to the TCP, displaying 
the same exponent as in the case of $\chi^{zz}$. 
A crucial difference from \fig{MFresult}(c) 
is that while 
the specific heat exhibits the same singularities even for $h=0$ 
if the PI is tuned, e.g., by the chemical potential \cite{yamase05}, 
the jump of $\chi^{zz}$ at a CP and its divergence at a TCP occur only for 
$h\neq 0$. This feature can be checked by observing that the second term in \eq{chizz} 
vanishes for $h=0$. 
Therefore the singular behavior of $\chi^{zz}$ is a field-induced anomaly. 
As we shall show below, it is a manifestation of a general feature 
of a non-ordering susceptibility and hence the presence of 
a finite non-ordering field (a magnetic field here) is necessary. 

\subsection{Scaling analysis and criticality beyond the RPA} 
So far we have considered the forward scattering model. Its universal features in the vicinity of CP and TCP are well described by mean-field theory, and are correctly captured also by a Landau-type $\phi^6$ theory \cite{lawrie84}.  
By allowing a small momentum transfer in \eq{fkk}, we can incorporate 
order parameter fluctuations. 
The obtained singularities of $\chi^{zz}$ and $C/T$ are expected to be modified 
in a critical regime. By virtue of universality, the character of the non-analyticities and the exact values of critical exponents can be deduced without performing explicit calculations. 
The order parameter describing the PI [\eq{ord_par_def}] is characterized by the Ising symmetry. 
The exact critical exponents of the Ising universality class are well 
known for ordering quantities. 
The clue to obtain the exact exponents of non-ordering quantities 
is seen in theory of tricritical points \cite{lawrie84}.  
The scaling hypothesis allows one to write the singular part of the 
Gibbs potential around a TCP as 
\be
G_{\rm sing}(t,g,s)=|g|^{2-\alpha}\mathcal{G}\left(t/|g|^{\phi}, s/|g|^{\Delta} \right),
\ee  
where $t=(T-T_{\rm tri})/T_{\rm tri}$ and $s$ is an ordering field which is 
conjugate to the order parameter; 
the scaling field $g$ is chosen as 
$g=(h-h_{\rm tri}) - at$ where $h$ is a non-ordering field and 
$a^{-1}$ is a slope of the transition line at the TCP. 
We focus on a situation where the transition line is approached with a finite angle.

Singular contributions to non-ordering quantities are obtained by taking 
derivatives of $G_{\rm sing}$ 
with respect to $h$ at $h=h_{\rm tri}$ and $s=0$ with $t\to 0$: 
\be
m=-\frac{\partial G_{\rm sing}}{\partial h}\sim |t|^{\beta_{2}}
\ee  
with $\beta_{2}=1-\alpha$ and 
\be
\chi=-\frac{\partial^{2} G_{\rm sing}}{\partial h^{2}}\sim |t|^{-\gamma_{2}}
\ee 
with $\gamma_{2}=\alpha$, where the additional subscript "2" indicates 
the exponents corresponding to non-ordering quantities. 
In the present case, the non-ordering field $h$ corresponds to a magnetic field and 
thus $m$ $(\chi)$ is 
the magnetization (longitudinal magnetic susceptibility).  
Since the specific heat coefficient $C/T$ at $h=h_{\rm tri}$ is given by 
\be
C/T=-\frac{\partial^{2} G_{\rm sing}}{\partial T^{2}}\sim |t|^{-\alpha},
\ee 
$\gamma_{2}$ is the 
same as the specific heat exponent. Hence we obtain the following scaling laws:  
\be
\alpha+ 2\beta_{2} +\gamma_{2}=2, \; \alpha=\gamma_{2}  \;.
\label{non-scaling}
\ee
For an ordinary CP,  replacing a TCP with a CP and employing the standard form 
\be
G_{\rm sing}(g,s)=|g|^{2-\alpha}\mathcal{G}\left(s/|g|^{\Delta} \right),
\ee 
we obtain the same scaling laws. Therefore \eq{non-scaling} holds for both CP and TCP. This is worth emphasizing that even the existence of TCP is not necessary for the fulfillment of the relation \eq{non-scaling}. 

\begin{table}[tb]
\begin{center}
\begin{tabular}{c|c|c|c|c}
& \multicolumn{2}{c|} {CP} &  \multicolumn{2}{c} {TCP} \\
\cline{2-5}
& {Mean field} & {\;Exact\; } & {Mean field} & {\;Exact\;} \\
\hline
$\alpha$ & 0 (jump) &  0 (log) & 1/2 & 8/9 \\
$\beta_{2}$ & 1 &  1 & 1/2 & 1/9 \\
$\gamma_{2}$ & 0 (jump) & 0 (log) & 1/2 & 8/9 \\
$\beta$ & 1/2 &  1/8 & 1/4 & 1/24 \\
$\gamma$ & 1 &  7/4 & 1 & 37/36 
\end{tabular}
\end{center}
\caption{Critical exponents at a CP and a TCP.  Mean-field values 
at a TCP are exponents on the symmetry-broken side. 
The exponents of ordering quantities are also listed for completeness: 
$\eta \sim |t|^{\beta}$  and $\kappa_{d}\sim |t|^{-\gamma}$, 
where in the present case 
$\eta$ corresponds to the order parameter of the PI [\eq{ord_par_def}] and $\kappa_{d}$ 
to the $d$-wave  compressibility \cite{yamase05}. 
}
\label{table}
\end{table}

Since the exact value of $\alpha$ is known both at CP and TCP \cite{lawrie84}, 
\eq{non-scaling} gives the exact values of $\beta_{2}$ and $\gamma_{2}$ (Table~\ref{table}). 
That is, $\chi^{zz}$ exhibits a divergence in a critical 
regime at both CP and TCP, indicating that the PI in a magnetic field is 
accompanied not only by soft Fermi surface fluctuations but also by 
sizable ferromagnetic fluctuations. 
It should be noted that the singularity of $\chi^{\pm}$ seen in \fig{MFresult} (d) 
is expected to remain the same even in the presence of fluctuations, 
because the relation $\chi^{\pm}=2m/h$ [\eq{chipm}] still holds beyond the RPA  
as long as the effective interaction between electrons retains the 
spin-rotational invariance \cite{matsumoto78},  
and $m/h$ is continuous at both CP and TCP.

\section{Discussion}
\subsection{Relevance to Sr$_{3}$Ru$_{2}$O$_{7}$} 
There is growing evidence that 
the bilayer ruthenate Sr$_{3}$Ru$_{2}$O$_{7}$ exhibits 
the PI \cite{borzi07,rost09,grigera04}. 
Although other interactions, not included in our model (1), may exist in Sr$_{3}$Ru$_{2}$O$_{7}$, 
by virtue of universality, the present approach is sufficient  
to correctly capture the universal singularities  (Table~\ref{table}) associated with the PI. 
Inclusion of other terms like spin-orbit coupling does not influence the order parameter
symmetry [\eq{ord_par_def}] and therefore the system can be presumed to remain in the 
two-dimensional Ising universality class even if the microscopic model is tailored to
capture more specific features of Sr$_3$Ru$_2$O$_7$.

The specific heat jump was reported recently \cite{rost09} 
at a CP of the PI. The actual critical region seems tiny 
enough not to be resolved in the experiment and the  observed jump 
corresponds to the mean-field behavior. 
Therefore we predict the same behavior also for 
$\chi^{zz}$ [Figs.3 (a) and (c)]. 
To the best of our knowledge, Sr$_{3}$Ru$_{2}$O$_{7}$ 
can provide the first example of a jump of the magnetic 
susceptibility at a continuous phase transition in condensed matter. 
On the other hand, at a TCP we predict a divergence of 
the jump of $\chi^{zz}$ [Figs.~3(b) and (c)] as well as 
that of $C/T$ (\fig{gamma}), although the latter becomes visible only sufficiently 
close to a TCP,  which requires high accuracy of an experimental measurement. 
The quantity $\chi^{\pm}$ is not a non-ordering susceptibility and 
its singularity might depend on microscopic details of the system. 
In particular, in Sr$_{3}$Ru$_{2}$O$_{7}$  the spin-rotational 
invariance is broken due to the spin-orbit coupling. 
Nonetheless, 
as far as the underlying interaction driving the PI [\eq{fkk}] does not  contribute 
to $\chi^{\pm}$ (\fig{diagram}), we expect 
a pronounced difference between $\chi^{zz}$ and $\chi^{\pm}$ at the PI. 

\subsection{Implications for other materials} 
As seen from the derivation of \eq{non-scaling}, the singular behavior of $\chi^{zz}$ is 
not a specific feature of the PI in the presence of a magnetic field, but 
can be interpreted as a phenomenon 
associated with singularities of a non-ordering susceptibility at a phase transition in general. 
This observation provides implications for a variety of materials. 
In particular, when a critical region is accessed by experiments, 
the mean-field predictions break down. 
In such a region, both ordering and non-ordering susceptibilities 
diverge simultaneously ($\gamma$ and $\gamma_{2}$ in Table~\ref{table}). 
The system features multiple critical fluctuations. 
The effect around a TCP may be more pronounced because the value of 
$\gamma_{2}$ is close to $\gamma$. 
Though the universality class is different from the present two-dimensional Ising type, 
singular behavior of a non-ordering susceptibility was indeed 
measured at a TCP in some metamagnetic compounds such as 
CsCoCl$_{3}\cdot$2D$_{2}$O and Dy$_{3}$Al$_{5}$O$_{12}$ 
in 1970s \cite{stryjewski77}.    
A manifestation of multiple singular susceptibilities is seen also in 
recent literature. For example in Ref.~\onlinecite{misawa0809} 
peculiar behavior of the uniform magnetic susceptibility was invoked to propose an explanation of anomalous properties of YbRh$_{2}$Si$_{2}$ and other antiferromagnetic compounds. 
Besides these materials, singular behavior of a non-ordering 
susceptibility, which has not been considered much in the modern condensed matter field, 
may provide an interesting route to interpret anomalous 
behavior observed in correlated electron systems. 

\section{Summary}
We have studied the uniform magnetic susceptibility at the PI on a square lattice 
in the presence of a magnetic field. 
In a mean-field model, $\chi^{zz}$ is found to exhibit a jump at a critical point 
and a power-law divergence at a tricritical point,  
whereas $\chi^{\pm}$ exhibits a cusp. 
The mean-field results may be modified in a critical region. 
In particular, $\chi^{zz}$ is expected to diverge also at a critical point, 
indicating that the PI in a magnetic field is accompanied not only by 
soft Fermi surface fluctuations but also by sizable ferromagnetic fluctuations. 
Invoking the scaling hypothesis we have argued that the features of $\chi^{zz}$ 
are not specific to the PI in a magnetic field, but are general properties 
associated with singularities of a non-ordering susceptibility at a phase transition. 
While such properties have 
not been recognized well in modern condensed matter literature, 
it can be tested for Sr$_{3}$Ru$_{2}$O$_{7}$ by measuring the longitudinal magnetic susceptibility.

\begin{acknowledgments}
We would like to acknowledge valuable discussions with A. A. Katanin, H. Kohno, W. Metzner, 
K. Miyake, and M. Napi\'{o}rkowski, and warm hospitality of the Max-Planck-Institute 
for Solid State Research in Stuttgart. 
\end{acknowledgments}


\bibliography{main.bib}

\begin{thebibliography}{30}
\expandafter\ifx\csname natexlab\endcsname\relax\def\natexlab#1{#1}\fi
\expandafter\ifx\csname bibnamefont\endcsname\relax
  \def\bibnamefont#1{#1}\fi
\expandafter\ifx\csname bibfnamefont\endcsname\relax
  \def\bibfnamefont#1{#1}\fi
\expandafter\ifx\csname citenamefont\endcsname\relax
  \def\citenamefont#1{#1}\fi
\expandafter\ifx\csname url\endcsname\relax
  \def\url#1{\texttt{#1}}\fi
\expandafter\ifx\csname urlprefix\endcsname\relax\def\urlprefix{URL }\fi
\providecommand{\bibinfo}[2]{#2}
\providecommand{\eprint}[2][]{\url{#2}}

\bibitem[{\citenamefont{de~Gennes and Prost}(1993)}]{deGennes93}
\bibinfo{author}{\bibfnamefont{P.~G.} \bibnamefont{de~Gennes}}
  \bibnamefont{and} \bibinfo{author}{\bibfnamefont{J.}~\bibnamefont{Prost}},
  \emph{\bibinfo{title}{The Physics of Liquid Crystals}}
  (\bibinfo{publisher}{Clarendon Press}, \bibinfo{address}{Oxford, UK},
  \bibinfo{year}{1993}).

\bibitem[{\citenamefont{Kivelson et~al.}(1998)\citenamefont{Kivelson, Fradkin,
  and Emery}}]{kivelson98}
\bibinfo{author}{\bibfnamefont{S.~A.} \bibnamefont{Kivelson}},
  \bibinfo{author}{\bibfnamefont{E.}~\bibnamefont{Fradkin}}, \bibnamefont{and}
  \bibinfo{author}{\bibfnamefont{V.~J.} \bibnamefont{Emery}},
  \bibinfo{journal}{Nature (London)} \textbf{\bibinfo{volume}{393}},
  \bibinfo{pages}{550} (\bibinfo{year}{1998}).

\bibitem[{yam()}]{yamase00}
\bibinfo{note}{H. Yamase and H. Kohno, J.\ Phys.\ Soc.\ Jpn.\ {\bf 69}, 332
  (2000); {\bf 69}, 2151 (2000).}

\bibitem[{\citenamefont{Halboth and Metzner}(2000)}]{metzner00}
\bibinfo{author}{\bibfnamefont{C.~J.} \bibnamefont{Halboth}} \bibnamefont{and}
  \bibinfo{author}{\bibfnamefont{W.}~\bibnamefont{Metzner}},
  \bibinfo{journal}{Phys.\ Rev.\ Lett.} \textbf{\bibinfo{volume}{85}},
  \bibinfo{pages}{5162} (\bibinfo{year}{2000}).

\bibitem[{\citenamefont{Valenzuela and Vozmediano}(2001)}]{valenzuela01}
\bibinfo{author}{\bibfnamefont{B.}~\bibnamefont{Valenzuela}} \bibnamefont{and}
  \bibinfo{author}{\bibfnamefont{M.~A.~H.} \bibnamefont{Vozmediano}},
  \bibinfo{journal}{Phys.\ Rev.\ B} \textbf{\bibinfo{volume}{63}},
  \bibinfo{pages}{153103} (\bibinfo{year}{2001}).

\bibitem[{\citenamefont{Pomeranchuk}(1958)}]{pomeranchuk58}
\bibinfo{author}{\bibfnamefont{I.~J.} \bibnamefont{Pomeranchuk}},
  \bibinfo{journal}{Sov.\ Phys.\ JETP} \textbf{\bibinfo{volume}{8}},
  \bibinfo{pages}{361} (\bibinfo{year}{1958}).

\bibitem[{wex()}]{wexler06}
\bibinfo{note}{C. Wexler and O. Ciftja, Int. J. Mod. Phys. B {\bf 20}, 747
  (2006) and references therein.}

\bibitem[{\citenamefont{{V. Hinkov, D. Haug, B. Fauqu\'{e}, P. Bourges, Y.
  Sidis, A. Ivanov, C. Bernhard, C. T. Lin, and B. Keimer}}(2008)}]{hinkov08}
\bibinfo{author}{\bibnamefont{{V. Hinkov, D. Haug, B. Fauqu\'{e}, P. Bourges,
  Y. Sidis, A. Ivanov, C. Bernhard, C. T. Lin, and B. Keimer}}},
  \bibinfo{journal}{Science} \textbf{\bibinfo{volume}{319}},
  \bibinfo{pages}{597} (\bibinfo{year}{2008}).

\bibitem[{\citenamefont{{R. A. Borzi, S. A. Grigera, J. Farrell, R. S. Perry,
  S. J. S. Lister, S. L. Lee, D. A. Tennant, Y. Maeno, and A. P.
  Mackenzie}}(2007)}]{borzi07}
\bibinfo{author}{\bibnamefont{{R. A. Borzi, S. A. Grigera, J. Farrell, R. S.
  Perry, S. J. S. Lister, S. L. Lee, D. A. Tennant, Y. Maeno, and A. P.
  Mackenzie}}}, \bibinfo{journal}{Science} \textbf{\bibinfo{volume}{315}},
  \bibinfo{pages}{214} (\bibinfo{year}{2007}).

\bibitem[{\citenamefont{{A. W. Rost, R. S. Perry, J.-F. Mercure, A. P.
  Mackenzie, and S. A. Grigera}}(2009)}]{rost09}
\bibinfo{author}{\bibnamefont{{A. W. Rost, R. S. Perry, J.-F. Mercure, A. P.
  Mackenzie, and S. A. Grigera}}}, \bibinfo{journal}{Science}
  \textbf{\bibinfo{volume}{325}}, \bibinfo{pages}{1360} (\bibinfo{year}{2009}).

\bibitem[{\citenamefont{{S. A. Grigera, P. Gegenwart, R. A. Borzi, F. Weickert,
  A. J. Schofield, R. S. Perry, T. Tayama, T. Sakakibara, Y. Maeno, A. G.
  Green, and A. P. Mackenzie}}(2004)}]{grigera04}
\bibinfo{author}{\bibnamefont{{S. A. Grigera, P. Gegenwart, R. A. Borzi, F.
  Weickert, A. J. Schofield, R. S. Perry, T. Tayama, T. Sakakibara, Y. Maeno,
  A. G. Green, and A. P. Mackenzie}}}, \bibinfo{journal}{Science}
  \textbf{\bibinfo{volume}{306}}, \bibinfo{pages}{1154} (\bibinfo{year}{2004}).

\bibitem[{\citenamefont{Lawrie and Sarbach}(1984)}]{lawrie84}
\bibinfo{author}{\bibfnamefont{I.~D.} \bibnamefont{Lawrie}} \bibnamefont{and}
  \bibinfo{author}{\bibfnamefont{S.}~\bibnamefont{Sarbach}}, in
  \emph{\bibinfo{booktitle}{Phase Transitions and Critical Phenomena vol.9}},
  edited by \bibinfo{editor}{\bibfnamefont{C.}~\bibnamefont{Domb}}
  \bibnamefont{and} \bibinfo{editor}{\bibfnamefont{J.~L.}
  \bibnamefont{Lebowitz}} (\bibinfo{publisher}{Academic Press},
  \bibinfo{address}{London}, \bibinfo{year}{1984}).

\bibitem[{\citenamefont{{E. Stryjewski and N. Giordano}}(1977)}]{stryjewski77}
\bibinfo{author}{\bibnamefont{{E. Stryjewski and N. Giordano}}},
  \bibinfo{journal}{Adv. Phys.} \textbf{\bibinfo{volume}{26}},
  \bibinfo{pages}{487} (\bibinfo{year}{1977}).

\bibitem[{gri()}]{griffiths7073}
\bibinfo{note}{R. B. Griffiths, Phys.\ Rev.\ Lett {\bf 24}, 715 (1970); Phys.\
  Rev.\ B {\bf 7}, 545 (1973)}.

\bibitem[{\citenamefont{{E. K. Riedel}}(1972)}]{riedel72}
\bibinfo{author}{\bibnamefont{{E. K. Riedel}}}, \bibinfo{journal}{Phys.\ Rev.\
  Lett.} \textbf{\bibinfo{volume}{13}}, \bibinfo{pages}{675}
  (\bibinfo{year}{1972}).

\bibitem[{\citenamefont{Yamase et~al.}(2005)\citenamefont{Yamase, Oganesyan,
  and Metzner}}]{yamase05}
\bibinfo{author}{\bibfnamefont{H.}~\bibnamefont{Yamase}},
  \bibinfo{author}{\bibfnamefont{V.}~\bibnamefont{Oganesyan}},
  \bibnamefont{and} \bibinfo{author}{\bibfnamefont{W.}~\bibnamefont{Metzner}},
  \bibinfo{journal}{Phys.\ Rev.\ B} \textbf{\bibinfo{volume}{72}},
  \bibinfo{pages}{035114} (\bibinfo{year}{2005}).

\bibitem[{\citenamefont{Yamase and Katanin}(2007)}]{yamase07b}
\bibinfo{author}{\bibfnamefont{H.}~\bibnamefont{Yamase}} \bibnamefont{and}
  \bibinfo{author}{\bibfnamefont{A.~A.} \bibnamefont{Katanin}},
  \bibinfo{journal}{J.\ Phys.\ Soc.\ Jpn.} \textbf{\bibinfo{volume}{76}},
  \bibinfo{pages}{073706} (\bibinfo{year}{2007}).

\bibitem[{\citenamefont{Yamase}(2007)}]{yamase07c}
\bibinfo{author}{\bibfnamefont{H.}~\bibnamefont{Yamase}},
  \bibinfo{journal}{Phys.\ Rev.\ B} \textbf{\bibinfo{volume}{76}},
  \bibinfo{pages}{155117} (\bibinfo{year}{2007}).

\bibitem[{\citenamefont{Kee and Kim}(2005)}]{kee05}
\bibinfo{author}{\bibfnamefont{H.-Y.} \bibnamefont{Kee}} \bibnamefont{and}
  \bibinfo{author}{\bibfnamefont{Y.~B.} \bibnamefont{Kim}},
  \bibinfo{journal}{Phys.\ Rev.\ B} \textbf{\bibinfo{volume}{71}},
  \bibinfo{pages}{184402} (\bibinfo{year}{2005}).

\bibitem[{\citenamefont{Doh et~al.}(2007)\citenamefont{Doh, Kim, and
  Ahn}}]{doh07}
\bibinfo{author}{\bibfnamefont{H.}~\bibnamefont{Doh}},
  \bibinfo{author}{\bibfnamefont{Y.~B.} \bibnamefont{Kim}}, \bibnamefont{and}
  \bibinfo{author}{\bibfnamefont{K.~H.} \bibnamefont{Ahn}},
  \bibinfo{journal}{Phys.\ Rev.\ Lett.} \textbf{\bibinfo{volume}{98}},
  \bibinfo{pages}{126407} (\bibinfo{year}{2007}).

\bibitem[{\citenamefont{{C. Puetter, H. Doh, and H.-Y. Kee}}(2007)}]{puetter07}
\bibinfo{author}{\bibnamefont{{C. Puetter, H. Doh, and H.-Y. Kee}}},
  \bibinfo{journal}{Phys.\ Rev.\ B} \textbf{\bibinfo{volume}{76}},
  \bibinfo{pages}{235112} (\bibinfo{year}{2007}).

\bibitem[{\citenamefont{Doh and Kee}(2007)}]{doh07b}
\bibinfo{author}{\bibfnamefont{H.}~\bibnamefont{Doh}} \bibnamefont{and}
  \bibinfo{author}{\bibfnamefont{H.-K.} \bibnamefont{Kee}},
  \bibinfo{journal}{Phys.\ Rev.\ B} \textbf{\bibinfo{volume}{75}},
  \bibinfo{pages}{233102} (\bibinfo{year}{2007}).

\bibitem[{\citenamefont{Ho and Schofield}(2008)}]{ho08}
\bibinfo{author}{\bibfnamefont{A.~F.} \bibnamefont{Ho}} \bibnamefont{and}
  \bibinfo{author}{\bibfnamefont{A.~J.} \bibnamefont{Schofield}},
  \bibinfo{journal}{Europhys. Lett.} \textbf{\bibinfo{volume}{84}},
  \bibinfo{pages}{27007} (\bibinfo{year}{2008}).

\bibitem[{\citenamefont{Yamase}(2009)}]{yamase09b}
\bibinfo{author}{\bibfnamefont{H.}~\bibnamefont{Yamase}},
  \bibinfo{journal}{Phys. Rev. B} \textbf{\bibinfo{volume}{80}},
  \bibinfo{pages}{115102} (\bibinfo{year}{2009}).

\bibitem[{\citenamefont{Adachi and Sigrist}(2009)}]{adachi09}
\bibinfo{author}{\bibfnamefont{H.}~\bibnamefont{Adachi}} \bibnamefont{and}
  \bibinfo{author}{\bibfnamefont{M.}~\bibnamefont{Sigrist}},
  \bibinfo{journal}{Phys.\ Rev.\ B} \textbf{\bibinfo{volume}{80}},
  \bibinfo{pages}{155123} (\bibinfo{year}{2009}).

\bibitem[{\citenamefont{{M. H. Fischer and M. Sigrist}}(2010)}]{fischer10}
\bibinfo{author}{\bibnamefont{{M. H. Fischer and M. Sigrist}}},
  \bibinfo{journal}{Phys. Rev. B} \textbf{\bibinfo{volume}{81}},
  \bibinfo{pages}{064435} (\bibinfo{year}{2010}).

\bibitem[{mis({\natexlab{a}})}]{miscchizz}
\bibinfo{note}{If the temperature instead of the field $h$ is used as a
  parameter to control a distance from the TCP, $\chi^{zz}\propto |h-h_{\rm
  tri}|^{-1/2}$ is obtained.}

\bibitem[{\citenamefont{{M. Blume, V. J. Emery, and R. B.
  Griffiths}}(1971)}]{blume71}
\bibinfo{author}{\bibnamefont{{M. Blume, V. J. Emery, and R. B. Griffiths}}},
  \bibinfo{journal}{Phys. \ Rev.\ A} \textbf{\bibinfo{volume}{4}},
  \bibinfo{pages}{1071} (\bibinfo{year}{1971}).

\bibitem[{\citenamefont{{H. Matsumoto, H. Umezawa, S. Seki, and M.
  Tachiki}}(1978)}]{matsumoto78}
\bibinfo{author}{\bibnamefont{{H. Matsumoto, H. Umezawa, S. Seki, and M.
  Tachiki}}}, \bibinfo{journal}{Phys.\ Rev.\ B} \textbf{\bibinfo{volume}{17}},
  \bibinfo{pages}{2276} (\bibinfo{year}{1978}).

\bibitem[{mis({\natexlab{b}})}]{misawa0809}
\bibinfo{note}{T. Misawa, Y. Yamaji, and M. Imada, J. Phys. Soc. Jpn. {\bf 77},
  093712 (2008); {\bf 78}, 084707 (2009)}.

\end{thebibliography}

\end{document}